# Optomechanics of liquid crystals for dynamical optical response of photonic structures

A. E. Miroshnichenko,[1] E. Brasselet,[2] D. O. Krimer,[3] and Yu. S. Kivshar[3]

[1]*Nonlinear Physics Centre and Centre for Ultra-high bandwidth Devices for Optical Systems (CUDOS), Australian National University, Canberra ACT 0200, Australia*
[2]*Centre de Physique Optique Moléculaire et Hertzienne, Université Bordeaux 1, CNRS, 351 Cours de la Libération, 33405 Talence Cedex, France*
[3]*Max Planck Institute for the Physics of Complex Systems, Nöthnitzer Str. 38, D-01187 Dresden, Germany*

We show that the mechanical effect of light on the orientational ordering of the crystalline axis of a mesophase can be used to control the dynamics of the optical response of liquid crystal infiltrated photonic structures. The demonstration is made using a one-dimensional periodic structure whose periodicity is broken by the presence of a nematic liquid crystal defect layer. In this study we report on output light polarization and/or intensity dynamics that depends on the initial molecular ordering and incident light wavelength and intensity.

## I. INTRODUCTION

Two decades ago, Yablonovitch [1] and John [2] independently introduced the concept of photonic crystals, i.e. dielectric systems with spatial periodicity of the refractive index, which leads to the formation of photonic bandgaps. The latter correspond to wavevector ranges (hence implying direction and frequency), for which light cannot propagate. Indeed, this feature allows to control the flow of light [3] provided an appropriate structured material properties. To do so, the light itself offers practical solutions towards high-resolution three-dimensional optical micro- and nano-fabrication techniques [4]. Although being a necessary first step, structured material is however not enough when flexible optical data processing is envisaged, which requires reconfigurable photonic crystals. To do so, linear or nonlinear refractive index changes have appeared as a straightforward solution that can be supplied by external fields of different nature (e.g., thermal, electrical, or optical). Obviously, the use of light itself is appealing since it naturally benefits from the strong spatial confinement of the field inside the structure, thereby enhancing the optical response and paving the way towards all-optical photonic circuitry [5].

Among various nonlinear optical materials, the attractiveness of liquid crystals result from the genuine combination of its 'liquid' and 'crystalline' features, which allow easy integration into photonic crystal microarchitectures and extreme sensitivity to external fields at the same time. Such a possibility has been thoroughly studied since the pioneering results of, on the one hand, Busch and John, who proposed to use the orientational ordering of the optical axis of liquid crystals in order to tune the properties of photonic bandgaps [6], and, on the other hand, Yoshino and coworkers, who proposed to benefit from the successive appearance of distinct mesophases in thermotropic liquid crystals when temperature is changed [7]. However, neither thermal nor electrical tuning of liquid crystals benefits from the very nature of a photonic crystal, contrary to the use of light itself to trigger refractive index changes.

The large orientational optical nonlinearities of liquid crystals [8, 9] can be fruitfully exploited to control the optical response of a photonic structure via the light that propagates inside it. Such an issue was addressed in previous works in the simplest case of a one-dimensional dielectric structure in which is embedded a pure or dye-doped nematic liquid crystal layer with initial uniform alignment [10–13]. These works, however, have only reported on static or transient light-induced changes. More recently, rotational dynamics of the director (i.e., the unit vector **n** that represents the local average molecular orientation of a nematic) has also been experimentally demonstrated [14].

In this study, we theoretically demonstrate that optomechanical effects at mesoscale can be used to control output light polarization and/or intensity dynamics of liquid crystal infiltrated photonic structures. First we introduce the system in section II A and the choice of two representative light-matter interaction geometries is discussed in section II B. Then, the optical threshold behavior is addressed in section III. The all-optical dynamical response of the photonic structures is studied in section IV and section V concludes the paper.

## II. BACKGROUND

### A. Definitions

We choose a one-dimensional periodic structure made of alternating layers of $SiO_2$ and $TiO_2$ with thicknesses 103 and 64 nm, respectively, which exhibits a bandgap in the visible range between 500 and 720 nm. In all our simulations, a uniformly aligned nematic liquid crystal defect layer with thickness $L = 5$ $\mu$m is located in the central part of the structure with five $SiO_2/TiO_2$ building blocks on each side (see Fig. 1).

We consider two different kind of initial uniform alignment $\mathbf{n}_0$ for the director. Namely, the initial alignment is either perpendicular (homeotropic anchoring, 'H') or parallel (planar anchoring, 'P') to the nematic slab. Hence, $\mathbf{n}_0 = \mathbf{e}_z$ or $\mathbf{n}_0 = \mathbf{e}_x$, where $(\mathbf{e}_x, \mathbf{e}_y, \mathbf{e}_z)$ is the Cartesian coordinate system, as illustrated in Fig. 1(a) and 1(b), respectively. We consider a nematic material with typical refractive indices $n_\perp = 1.5$ and $n_\parallel = 1.7$, where symbols $(\perp, \parallel)$ refer to directions perpendicular and parallel to $\mathbf{n}$, respectively, and Frank elastic constants are chosen such that $K_1/K_3 = 2/3$ and $K_2/K_3 = 1/2$.

The optical properties of these structures are calculated in the plane wave approximation, therefore all variables depend on coordinate $z$ and time $t$ only. The light propagation problem is solved by using the Berreman $4 \times 4$ matrix approach [15] and taking into account that, inside the nematic, the light is coupled to the Euler-Lagrange equations that govern the dynamics of the director [16]. We also introduce the reduced spatial coordinate $\xi = z/L$ and time $\tau = t/\tau_{H,P}$ where we defined the geometry-dependent characteristic times $\tau_H = \gamma_1 L^2/(\pi^2 K_3)$ [17, 18] and $\tau_P = \gamma_1 L^2/(\pi^2 K_2)$ [19], $\gamma_1$ being the rotational viscosity.

The director is represented by the usual spherical angles $\Theta$ and $\Phi$ following $\mathbf{n} = (\sin\Theta \cos\Phi, \sin\Theta \sin\Phi, \cos\Theta)$ but the different boundary conditions at $\xi = 0$ and $\xi = 1$ for the H and P geometries impose to distinguish these two cases. In the homeotropic case, $\mathbf{n}(0,\tau) = \mathbf{n}(1,\tau) = \mathbf{e}_z$ and the director representation is

$$\mathbf{n}_H = (\sin\Theta_H \cos\Phi_H, \sin\Theta_H \sin\Phi_H, \cos\Theta_H), \quad (1)$$

with

$$\Theta_H(\xi,\tau) = \sum_{m\geq 1} \Theta_H^{(m)}(\tau) \sin(m\pi\xi), \quad (2)$$

$$\Phi_H(\xi,\tau) = \Phi_H^{(0)}(\tau) + \sum_{m\geq 1} \Phi_H^{(m)}(\tau) \frac{\sin[(m+1)\pi\xi]}{\sin(\pi\xi)}, \quad (3)$$

$m$ integer, whereas in the planar case the director lies in the plane $(x,y)$ with $\mathbf{n}(0,\tau) = \mathbf{n}(1,\tau) = \mathbf{e}_x$, hence

$$\mathbf{n}_P = (\cos\Phi_P, \sin\Phi_P, 0), \quad (4)$$

with

$$\Phi_P(\xi,\tau) = \sum_{m\geq 1} \Phi_P^{(m)}(\tau) \sin(m\pi\xi). \quad (5)$$

In practice, we retain a large enough number of polar and azimuthal modes [i.e., sums that appear in Eqs. (2,3,5) are truncated] in order to ensure accurate results, which is done following previous works devoted to nematic slab alone in the case H [17, 18] and P [19]. We introduce the (adiabatic) total phase delay $\Delta$ between the extraordinary ($e$) and ordinary ($o$) waves due to the nematic layer (which turns out to be a measure of the reorientation associated with the polar degree od freedom of the director). Again, distinction must be made between H and P cases. In the homeotropic case

$$\Delta_H(\tau) = k_0 L \int_0^1 [n_{\text{eff}}(\xi,\tau) - n_\perp] d\xi, \quad (6)$$

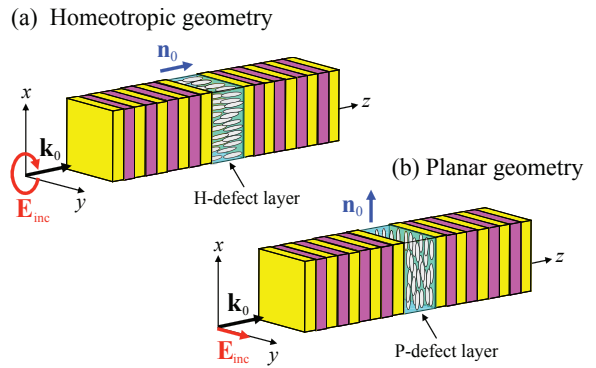

FIG. 1. (Color online) Definition of the homeotropic (a) and planar (b) geometries. They consist in a one-dimensional multilayered structure made of $SiO_2$ (yellow slabs) and $TiO_2$ (magenta slabs) in which is embedded a nematic liquid crystal layer. The uniform spatial distribution of its optical axis at rest is $\mathbf{n}_0 = \mathbf{e}_z$ in the case H and $\mathbf{n}_0 = \mathbf{e}_x$ in the case P. The excitation light beam impinges at normal incidence onto the structures, $\mathbf{k}_0 = k_0 \mathbf{e}_z$, and the incident polarization state is set to circular in the case H, $\mathbf{E}_{\text{inc}} = E_0(\mathbf{e}_x + i\mathbf{e}_y)/\sqrt{2}$, and to linear in the case P, $\mathbf{E}_{\text{inc}} = E_0 \mathbf{e}_y$.

with $n_{\text{eff}} = n_\parallel n_\perp/(n_\parallel^2 \cos^2\Theta_H + n_\perp^2 \sin^2\Theta_H)^{1/2}$, $k_0 = 2\pi/\lambda_0$ and $\lambda_0$ is incident wavelength in free space, whereas the condition $\Theta_P = \pi/2$ in the planar case imposes a constant phase delay

$$\Delta_P = k_0 L(n_\parallel - n_\perp). \quad (7)$$

Finally, we define the reduced incident light intensities as $\rho_H = I_{\text{inc}}/I_H$ in the homeotropic case, where $I_{\text{inc}}$ is the incident intensity and $I_H = 2\pi^2 c K_3 n_\parallel^2/[(n_\parallel^2 - n_\perp^2) n_\perp L^2]$ is the Fréedericksz threshold (i.e. the intensity above which the initial director state $\mathbf{n}_0$ is unstable) for circularly polarized light for a slab alone [16], whereas $\rho_P = I_{\text{inc}}/I_P$ with $I_P = 8\pi^2 c K_2 n_\parallel (n_\parallel - n_\perp)/[\lambda^2(n_\parallel + n_\perp)]$ in the planar case [19], where $c$ is the speed of light in free space.

### B. Light-matter interaction geometries

A general feature of the optical reordering of liquid crystals is its strong dependence on the light-matter interaction geometry, namely the director field at rest and the incidence angle, polarization and intensity of the incident light field. Therefore a preliminary analysis is performed for the H and P cases from the knowledge of the reorientation dynamics for a nematic slab alone (i.e., without periodic structure) in order to identify situations that will potentially exhibit a dynamical optical response.

*Homeotropic case* — The optical reordering of a homeotropic nematic film has been intensively studied for normal or oblique incidence and for a various set of incident beam polarization state, be it linear, elliptical, circular or unpolarized. Various static, periodic, quasi-periodic and aperiodic reorientation dynamics have been predicted and observed (see [8] for a review). The circular polarization case at normal incidence is a representative example for which periodic, quasi-periodic and chaotic rotations (see [20] and references therein) have all been reported. Therefore, it seems reasonable to expect dynamical behavior too when a periodic structure is at work. However, we note that a thin nematic slab can be a crippling drawback since the light-induced dynamical richness declines as the thickness decreases, as shown in Fig. 2. The phase delay $\Delta_\mathrm{H}$, and the collective director precession angular velocity, $\Omega = d\Phi_\mathrm{H}^{(0)}/d\tau$, are plotted in Fig. 2(a) and 2(b), respectively, for two different thicknesses. From this figure, we see that the precession-nutation regime (i.e., $\Omega \neq 0$ with $\partial\Delta_\mathrm{H}/\partial\tau \neq 0$) disappears below a typical cell thickness that corresponds to $L = 2-3$ $\mu$m for typical wavelength $\lambda_0 = 600$ nm. The typical reorientation dynamics diagram observed for thick cells ($L \sim 100$ $\mu$m, see [17]) is thus qualitatively preserved in our case, where $L = 5$ $\mu$m.

*Planar case* — The planar alignment geometry for nematics at normal incidence has been much less studied than its homeotropic counterpart. To our knowledge, this geometry has only been considered in [21] and no further studies have been performed on it during the past two decades. This might be explained by a (twist) Fréedericksz transition threshold predicted to be up to three orders of magnitude larger than the homeotropic case, thus preventing from experimental observation. Moreover, the above-threshold director distortions have been predicted to be static whatever the light intensity [21], hence limiting the need for a deeper analysis. Nevertheless, very recently, the planar geometry has attracted a renewal of interest following seemingly experimentally accessible intensity threshold using a periodic structure [12].

In fact, the case of a planar slab alone was revisited in [19] where it is rigorously demonstrated that the reorientation threshold is significantly lower than previously predicted [21]. Moreover, a static distorted state is expected above the reorientation threshold only for optically thin enough liquid crystal layers, $\Delta_\mathrm{P}/\pi < 0.64$ [19], whereas a dynamical regime that takes place via a Hopf bifurcation is predicted otherwise, which was missed so far [21]. A secondary instability (heteroclinic bifurcation) has also been predicted, which eventually leads to a static regime, is associated ($0.64 < \Delta_\mathrm{P}/\pi < 1$), or not ($\Delta_\mathrm{P}/\pi > 1$), to a hysteresis behavior [19]. These results are summarized in Fig. 3 where the reduced third component of the Stokes parameter at the output is plotted as a function of the reduced input intensity $\rho_\mathrm{P}$ for three representative optical thicknesses that correspond to $\Delta_\mathrm{P}/\pi = 0.5$ [panel(a)], 0.8 [panel(b)] and 1.6 [panel(c)].

Since $L = 5$ $\mu$m corresponds to $\Delta_\mathrm{P}/\pi \sim 4$ in the visible range with the present choice of material parameters, a non trivial sequence of nonlinear reorientation dynamics is a priori expected. Importantly, note that the chosen film thickness is nevertheless thin enough to prevent from the influence of stimulated scattering, in contrast to thick cells where full polarization conversion can be observed [22, 23]. Indeed, the latter phenomenon is all the more important as the ratio $\Lambda = L(n_\parallel - n_\perp)/\lambda_0$ between the thickness and the intensity modulation grating period arising from the coherent superposition of extraordinary and ordinary waves is large compared to unity. Since $\Lambda \sim 2$ in our case, the stimulated scattering can therefore be safely neglected in our approach.

To summarize, we further consider the case H under circularly polarized excitation at normal incidence and the case P under ordinary linearly polarized excitation at normal incidence, as sketched in Figs. 1(a) and 1(b), respectively.

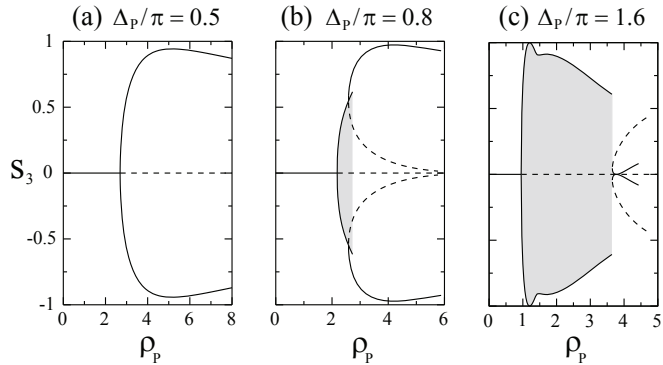

FIG. 3. P slab alone. Output reduced third Stokes parameter $s_3$ vs. intensity for three different nematic thicknesses that correspond to $\Delta_\mathrm{P}/\pi = 0.5$ (a), 0.8 (b) and 1.6 (c). Solid (dashed) curves refer to stable (unstable) states. The gray regions in panels (b) and (c) correspond to the range of values explored during the oscillatory regime.

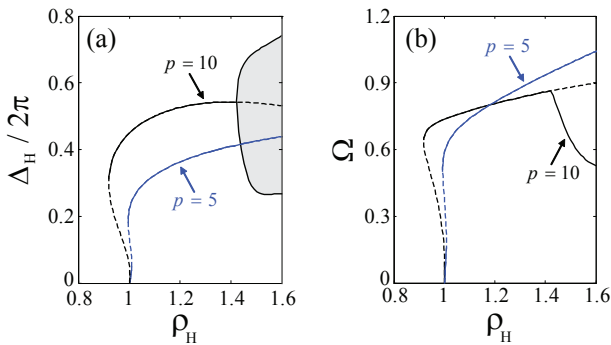

FIG. 2. (Color online) H slab alone. Total phase delay $\Delta_\mathrm{H}$ (a) and director precession angular velocity $\Omega = d\Phi_\mathrm{H}^{(0)}/d\tau$ (b) vs. intensity for two different nematic thicknesses $L = p\lambda$ ($p = 5$ and 10), where $\lambda = \lambda_0/n_\perp$ is the wavelength in the unperturbed nematic. Solid (dashed) curves refer to stable (unstable) states. The gray region in panel (a) corresponds to the range of values explored during the quasi-periodic regime.

## III. SPECTRALLY MODULATED OPTOMECHANICAL EFFICIENCY

In both H and P cases, the incident beam polarization corresponds to an ordinary wave for the nematic at rest and the optical torque density exerted on $\mathbf{n}_0$ is thus zero. However the orientational state $\mathbf{n} = \mathbf{n}_0$ is linearly unstable with respect to fluctuations when the light intensity is high enough to overcome the restoring elastic torque density that originates from the cell walls, which defines the optical Fréedericksz transition threshold [16]. The presence of the periodic structure does not suppress the optical destabilization scenario, but enriches it. Indeed, the reduced thresholds $\rho_{\rm H,th}$ [Fig. 4(a)] and $\rho_{\rm P,th}$ [Fig. 4(b)] strongly depends on the excitation beam wavelength $\lambda_0$ inside the bandgap, as already discussed in details in the homeotropic case [11, 13]. Namely, the defect modes evidenced by sharp peaks in the transmission spectrum $T_{\rm H}$ [Fig. 4(c)] are associated to an enhanced light confinement in the nematic defect layer that lead to a significant reduction (up to several orders of the magnitude) of the required incident intensity to trigger the liquid crystal reorientation [Fig. 4(a)].

In the planar case, a strongly modulated reorientation threshold is observed too, as shown in [Fig. 4(b)], however, its fine structure is more complex than its homeotropic counterpart. Indeed the spectrum of $\rho_{\rm P,th}$ exhibits two sets of resonances with either symmetric [circle symbols in Fig. 4(b)] or asymmetric [square symbols in Fig. 4(b)] lineshape that are related to the o- and e-defect mode wavelengths $\lambda_{\rm d}^o$ and $\lambda_{\rm d}^e$, respectively. Such a behavior is a direct manifestation of competing influence of the o- and e-modes. On the one hand, symmetric resonance lineshapes in the planar case are reminiscent of the incident ordinary polarized light field enhancement in the nematic defect layer when $\lambda_0 \approx \lambda_{\rm d}^o$, as in the homeotropic case. On the other hand, the origin of the asymmetric resonance lineshapes when $\lambda_0 \approx \lambda_{\rm d}^e$ can be qualitatively grasped by considering a fluctuation around the initial state $\mathbf{n}_0$ for a planar slab alone, see Fig. 5(a). The elastic torque density ($\mathbf{\Gamma}_{\rm el}$) driven by the cell walls tends to restore the director perturbation whereas the subsequent polarization changes (see typical polarization state inside the liquid crystal labeled as $\mathbf{E}_{\rm LC}$ in Fig. 5) are associated to a spin angular momentum deposition [24], hence an optical torque density ($\mathbf{\Gamma}_{\rm opt}$), that tends to increase the initial amplitude of the fluctuation, as illustrated in Fig. 5(a). The influence of the periodic structure is addressed by noting that (i) the phase of the o-wave is almost constant in the vicinity of a e-defect mode (except when $\lambda_{\rm d}^o \approx \lambda_{\rm d}^e$) and (ii) a $\pi$ phase shift is experienced by the e-wave when incident beam wavelength explores the resonance linewidth. Therefore, the sign of the optical feedback ($\delta\mathbf{\Gamma}_{\rm opt}$) is either negative or positive feedback depending on the resonance side, as shown in Fig. 5(b) and 5(c). We note here the analogy with the Fano resonance arising from the interference of resonant and non-resonant scattering waves, which is characterized by an asymmetric lineshape [25]. Indeed, resonantly excited e-wave leads to constructive and destructive interference with o-wave in the vicinity of a e-defect mode. When $\lambda_{\rm d}^o \approx \lambda_{\rm d}^e$ [i.e., $\lambda_0 \approx 673$ nm, see arrow in Fig. 4(d)], the dephasing associated to the defect modes between the o- and e-waves is even smaller than $\lambda_{\rm d}^o$ is close to $\lambda_{\rm d}^e$, which eventually leads to symmetric resonance lineshape for $\rho_{\rm P,th}$ [see arrow in Fig. 4(b)].

The nature of the optical reordering transition also depends on the excitation beam wavelength. In the homeotropic case, the order of the orientational instability can be either first- or second-order depending on the detuning $\lambda_0 - \lambda_{\rm d}^o$ between the pump wavelength and the nearest defect mode $\lambda_\mathrm{d}^o$, as previously discussed for identical structure with linearly polarized excitation [11, 13], which is a general result for homeotropic films inside a resonator under linearly polarized light [26, 27]. Indeed, the transition is first-order for positive detuning whereas it is second-order for negative detuning [see later Fig. 8(a) and 8(b) in the next section]. Such a behavior is therefore independent on the presence or not of an intrinsic optical bistability (i.e., for the nematic slab alone) since linearly and circularly polarized excitations are related to second- and first-order transitions, respectively. In the planar case, we also find that the order of the transition is controlled by the detuning with respect to the nearest $o$-defect mode, whatever the detuning between the $o$- and $e$-defect modes, but with some exception in the vicinity of the $e$-defect modes where a supercritical Hopf bifurcation is found (i.e., second-order transition), as illustrated in Fig. 6. This emphasizes the prominent role played by the $e$-defect modes in the case P although incident light is ordinary polarized.

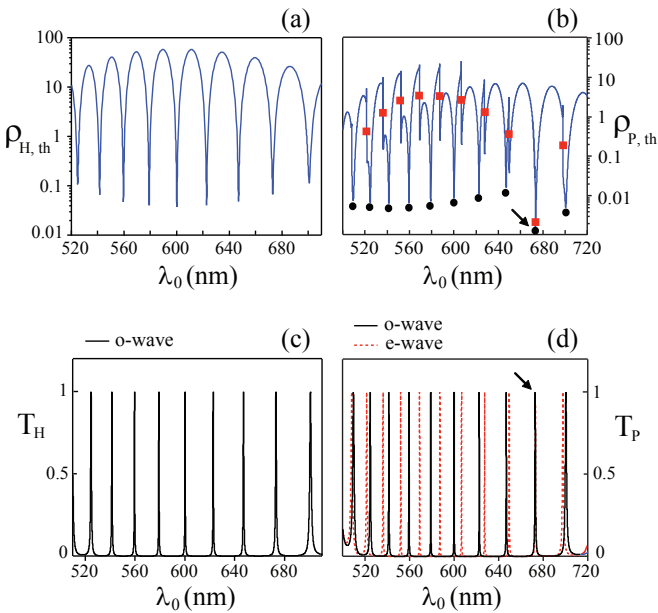

FIG. 4. (Color online) Reduced thresholds $\rho_{\rm H,th}$ (a) and $\rho_{\rm P,th}$ (b) and transmission spectra $T_{\rm H}$ (c) and $T_{\rm P}$ (d) at rest ($\mathbf{n} = \mathbf{n}_0$). In panel (b), circle (square) symbols refer to the ordinary (extraordinary) mode shown in panel (d).

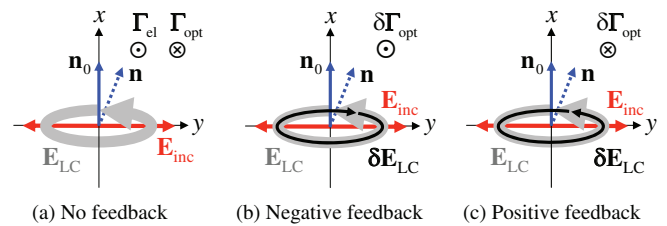

FIG. 5. (Color online) Qualitative interpretation of the asymmetric resonance lineshape nearby an extraordinary defect mode in the planar geometry. Elastic and optical torque densities exerted on the director in presence of an orientational fluctuation are shown in panel (a) for a slab alone whereas panels (b) and (c) illustrate the negative and positive feedback from the periodic structure for $\lambda_0 \lesssim \lambda_{\rm d}^e$ and $\lambda_0 \gtrsim \lambda_{\rm d}^e$, respectively.

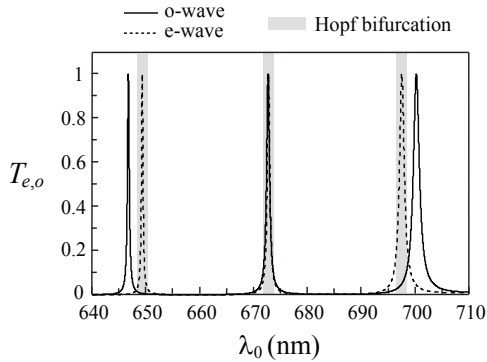

FIG. 6. Spectrally modulated nature of the reorientation transition that corresponds to a stationary bifurcation except in the vicinity of $e$-defect modes, which is identified by gray regions, where a Hopf bifurcation takes place.

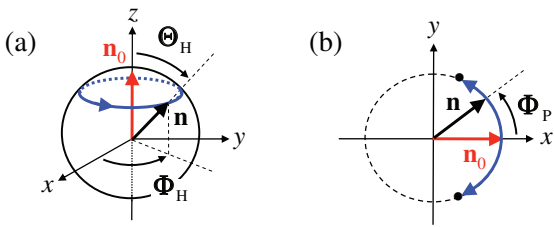

FIG. 7. (Color online) Director regimes in the H (a) and P (b) geometries. Uniform precession of the director around the $z$ axis is found in the case H whereas either static distorted states (see markers) or director oscillation in the $(x,y)$ plane are predicted in the case P.

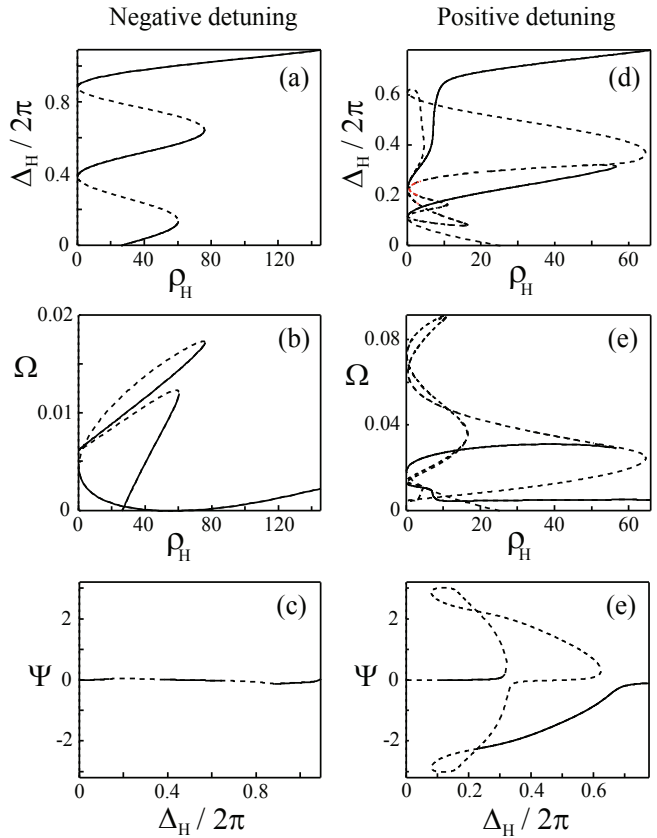

FIG. 8. Director reorientation characteristics for negative [panels (a,b,c)] and positive [panels (d,e,f)] detuning $\lambda_0 - \lambda_\mathrm{d}^o = \pm 5$ nm near the defect mode $\lambda_\mathrm{d}^o \approx 600$ nm. The total phase delay $\Delta_\mathrm{H}/2\pi$ and angular rotation frequency $\Omega$ vs. input intensity are shown in panels (a,d) and (b,e), respectively. The total twist defined by $\Psi = \Phi_\mathrm{H}(1,\tau) - \Phi_\mathrm{H}(0,\tau)$ vs. $\Delta_\mathrm{H}/2\pi$ is shown in panels (c,f). Solid (dashed) curves correspond to stable (unstable) regimes.

## IV. ALL-OPTICAL DYNAMICAL OPTICAL RESPONSE

As expected from the analysis of H and P nematic slab alone in section II B, we find dynamical regimes for the director reorientation above $\rho_\mathrm{H,th}$ and $\rho_\mathrm{P,th}$, respectively. In the case H, however, there is no nutation regime ($\partial \Theta_\mathrm{H}/\partial \tau \neq 0$) whatever is the wavelength although uniform precession ($\partial \Phi_\mathrm{H}/\partial \tau = \mathrm{constant}$) is always present. In the case P, the distorted state is static or correspond to an oscillatory motion in the $(x,y)$ plane. These regimes are sketched in Fig. 7 and next we discuss the optical response of the structure when dynamics takes place.

*Homeotropic case* — The uniform director nutation-free precession dynamics corresponds to a constant polar spatial profile ($\Theta_\mathrm{H}$) for which $\Delta_\mathrm{H}$ is an estimate of the overall amplitude [see Eq. (6)]. The azimuthal spatial profile is also constant but in a frame that rotates at constant angular velocity $\Omega = \partial \Phi_\mathrm{H}/\partial \tau = d\Phi_\mathrm{H}^{(0)}/d\tau$ around the $z$ axis. Also, we define the overall twisted charac-ter of the distortion profile in the rotating frame as the twist angle between the output and input facets of the nematic defect layer, $\Psi = \Phi_H(1,\tau) - \Phi_H(0,\tau)$. The incident intensity dependence of these three representative quantities, $\Delta_H$, $\Omega$ and $\Psi$ are shown in Fig. 7 for negative and positive detuning, $\lambda_0 - \lambda_d^o = \pm 5$nm, near the defect mode $\lambda_d^o \approx 600$ nm [see Fig. 4(c)].

In fact, $\Delta_H$ and $\Omega$ are related to distinct features of the polarization state at the output of the structure, namely the ellipticity and the rotation rate of the polarization ellipse, whose azimuth is given by $\Phi_H(1,\tau)$ (non-adiabatic light propagation when $\Psi \neq 0$ slightly modify this correspondence) [17, 24]. Since $\Delta_H$ is time-independent, the total transmission $T_H$ is constant and the dynamical optical response of the structure is merely contained in the polarization azimuth whatever the sign of the detuning. The intensity dependence of the optical response, however, strongly depends on wavelength, as shown in Fig. 8. Indeed, for negative detuning, the overall reorientation picture is very similar to the case of linearly polarized excitation [11], as shown in Fig. 8(a). This is due to the almost untwisted distortion profile, $\Phi_H^{(m)} \ll 1$, hence $\Psi \ll 1$, see Fig. 8(c). This is in stark distinction with the case of positive detuning where a complex reorientation diagram is found [Fig. 8(c)] as a result of large twist amplitude and associated significant non adiabatic light propagation effects. Such a behavior is not observed for a homeotropic slab alone and is the signature of the interaction between o- and e-defect modes that are no longer degenerate when $\mathbf{n} \neq \mathbf{n}_0$.

*Planar case* — Static or oscillatory regimes are found depending on the incident wavelength and intensity. Static distortions are found for almost any wavelength except in the vicinity of extraordinary defect modes where oscillatory dynamics appears via a Hopf bifurcation, as illustrated in Fig. 9(a). In that case, the output optical response exhibits time-dependent o- and e-transmissions, as shown in Fig. 9(b), however noting that the modula-tion depth of $T_o(\tau)$ are much less pronounced than the one of $T_e(\tau)$ and $T_e \ll T_o$. Details of the dynamics is illustrated in Fig. 10 near and well above the threshold at intensities that correspond to labels A and B in Fig. 9(a), respectively. As expected from a Hopf bifurcation, a single frequency characterizes the dynamics near $\rho_{P,\text{th}}$, see Fig. 10(a), whereas a second frequency grows up as the intensity increases, which leads to a quasi-periodic oscillation with the subsequent appearance of harmonic frequencies and frequency mixing that are clearly seen in the Fourier spectrum of the time-dependent transmission, see Fig. 10(b). The corresponding spatio-temporal director dynamics, $\Phi_P(\xi,\tau)$, is shown in Fig. 10(c,d). Aperiodic dynamics, which has not been predicted for a planar slab alone [19], is also found in simulations at higher intensities (not shown here), which emphasizes how the presence of the periodic structure contrary to the homeotropic case.

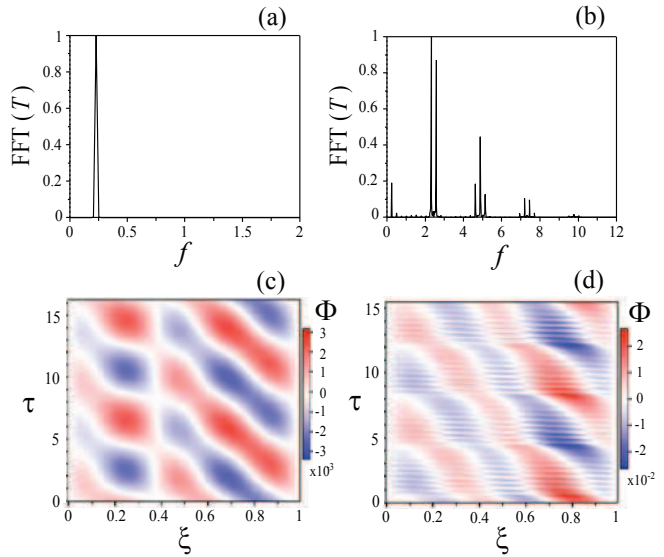

FIG. 10. (Color online) Fourier spectra of the time-dependent total transmission [panels (a) and (b)] and spatio-temporal dynamics of the azimuthal spatial profile $\Phi_P(\xi,\tau)$ [panels (c) and (d)] just above the Hopf bifurcation threshold [see A in Fig. 9(a)] and far above the threshold [see B in Fig. 9(a)].

## V. CONCLUSION

The dynamical response of all-optical liquid crystal infiltrated photonic structure has been discussed in two light-matter interaction geometries that correspond to a circularly [linearly] polarized incident beam impinging at normal incidence onto a one-dimensional periodic structure in which is embedded a nematic liquid crystal defect layer with homeotropic [planar] alignment (i.e., perpendicular [parallel] to the layer). In both cases the incident beam polarization state corresponds to an ordinary wave and optical reordering takes place only above a thresh-old for the input intensity that is much smaller than the usual optical Fréedericksz transition threshold reduced with respect to in the vicinity of ordinary defect modes. It has been shown that self-sustained dynamics of the output light polarization and/or intensity can take place in contrast with all previous attempts discussed so far, thus paving the way towards all-optical dynamical response of photonic structures using liquid crystals. For example this could be used to exploit dynamically the recently introduced concept of reversible optical nonreciprocity [28].

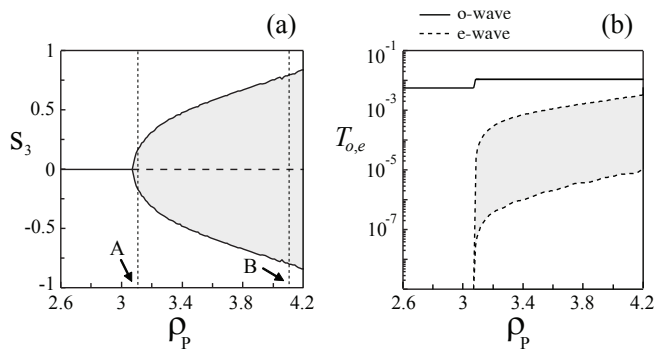

FIG. 9. (a) Reduced Stokes parameter $s_3$ vs. input intensity where solid (dashed) curves refer to stable (unstable) states. (b) Ordinary and extraordinary transmissions $T_o$ (solid curve) and $T_e$ (dashed curve) vs. input intensity at $\lambda_0 = 650$nm. Gray areas correspond to the range of values explored during the dynamics.

This work was supported by Australian Research Council through the Discovery Project and Centre of Excellence programs and by the FranceAustralia cooperation project 21337 of CNRS.